\documentclass{eptcs}

\usepackage{amsfonts,amssymb,latexsym,xspace}
\usepackage{amsmath,enumerate}
\usepackage[english]{babel}
\usepackage{graphicx}
\usepackage{amsfonts,latexsym}

\def \cal {\mathcal}

\def\tilde{\widetilde}
\def \inc { \subset^{\hskip -0.1cm\tilde {}}  \hskip 0.1cm}

\def\cal {\mathcal}

 \def \Pr {{\rm Pr}}

 \def\JF {}

 \def \pn {\par \noindent}
 \def\HH{{\cal H}}
  
  \def \H {{\bf H}}
  \def \bH  {{\mathbb H}}
  \def \bM {{\mathbb M}}

\newtheorem{Pro}{Proposition}
\newtheorem{Thm}{Theorem}
\newtheorem{Def}{Definition}

\title%[Information theory : Sources, Dirichlet series, and realistic analysis.]
 {Information theory : Sources, Dirichlet series,\\ and realistic analyses of data structures.
}
\author{
Mathieu Roux
\institute{LMNO  and  GREYC (CNRS and  University of Caen), France}
\email{mathieu.roux@unicaen.fr}
\and Brigitte Vall\'ee
\institute{ GREYC (CNRS and   University of Caen), France}
\email{brigitte.vallee@unicaen.fr}
}
\def\titlerunning{Information theory : Sources, Dirichlet series, and realistic analyses}
\def\authorrunning{M. Roux \& B. Vall\'ee}

\begin{document}

\maketitle

\begin{abstract} 
Most of the text algorithms build data structures on words, mainly trees, as digital  trees (tries) or binary search trees (bst). The mechanism which produces symbols of the words (one symbol at each unit time) is called a source, in information theory contexts.  The probabilistic behaviour of the trees built on words emitted by the same source depends on two factors: the  algorithmic properties of the tree, together with the information-theoretic properties of the source. Very often, these two factors are considered in a too simplified way: from the algorithmic point of view,  the cost of the Bst is only measured in terms of the number of comparisons between words --from the information theoretic point of view, only simple sources (memoryless sources or Markov chains)  are studied. \\
We wish to perform  here a realistic analysis, and we choose to deal  together with a general source and 
a realistic cost for data structures: we take into account comparisons between symbols, and we consider a general model of source, related to  a dynamical system, which is called  a dynamical source.    Our methods are close to analytic combinatorics, and our main object of interest is the generating function of the source $\Lambda(s)$, which is here of Dirichlet type. Such an object transforms probabilistic properties of the source into analytic properties. The tameness of the source, which is defined through analytic properties of $\Lambda(s)$,  appears to be central in the analysis, and is precisely studied for the class of dynamical sources. We focus here on arithmetical conditions, of diophantine type, which are sufficient to imply  tameness on a domain with hyperbolic shape. 

\end{abstract}

\paragraph{\bf Plan of the paper.} We first recall in Section 1 general facts on sources and trees, and define the probabilistic model chosen for the analysis.  Then, we provide the statements of the main two theorems  (Theorem 1 and 2) which establish the possible probabilistic behaviour of trees, provided that the source be tame. The tameness notions are  defined  in a general framework  and  then studied in the case  of simple sources (memoryless sources and Markov chains). In Section 2,  we  focus on   a general model of sources,  the dynamical sources, that contains as a subclass the simple sources.    We present sufficient conditions on  these sources under which it is possible to prove tameness. We  compare these tameness properties to those of simple sources, and exhibit both resemblances and differences between the two classes.

\section  {Probabilistic behaviour  of trees built on general sources.}

\medskip{\bf 1.1. General sources.}  Throughout this paper, an ordered (possibly infinite denumerable) alphabet $\Sigma:= \{a_1, a_2, \ldots, a_r \}$ is fixed.

A \emph{probabilistic   source},  which  produces  infinite  words  of
$\Sigma^{\mathbb N}$, is specified by the set $\{p_w, w \in \Sigma^\star\}$
of  \emph{fundamental probabilities}  $p_w$, where  $p_w$  is the probability
that an  infinite word begins with the  finite  prefix $w$.  It is furthermore assumed
that 
$\pi_k  := \sup \{p_w:w \in \Sigma^k \}$ tends \JF
to $0$, as $k \to \infty$.

  As it is usual in the  domain of analytic combinatorics, well described in  \cite{FlSe09}, our analyses 
involve the  generating function of the source,  here of Dirichlet type, first introduced in \cite {Vallee01} and defined as 
\begin{equation}\label{Lambdadef} 
\Lambda(s) := \sum_{w \in \Sigma^\star} p_w^{s},  \qquad \Lambda_{(k)} (s) := \sum_{w \in \Sigma^k} p_w^{s} .
\end{equation}
Since  all the equalities $\Lambda_{(k)}(1) = 1$ hold,  the series $\Lambda(s)$   is  divergent   
at $s = 1$, and the   probabilistic    properties of the source can be expressed in terms 
of   the regularity of $\Lambda$ near $s=1$, as it is known from previous works  \cite{Vallee01} and  will be recalled  later.  For instance, the entropy  $h({\cal S})$ relative to a  probabilistic
source ${\cal S}$ is defined as the limit (if it exists) 
that involves the  previous Dirichlet series \begin{equation}\label{entropy}
h({\cal S}):=\lim_{k\to \infty} \frac {-1} k \sum_{w \in \Sigma^k}p_w\log p_w =  \lim_{k \to \infty}   \frac {-1} k  \frac {d}{ds} \Lambda_{(k)} (s)_{|_{s = 1}}.
\end{equation}

\medskip {\bf 1.2. Simple sources:  memoryless sources and Markov chains.}
A memoryless source,  associated to the  (possibly infinite) alphabet $\Sigma$,   is defined by the  set  $(p_j)_{j \in \Sigma}$ of probabilities, and the Dirichlet series $\Lambda, \Lambda_{(k)}$  are expressed with
\begin{equation} \label{lambda1}
  \lambda (s) =  \sum_{i \in \Sigma} p_i^s,\qquad \hbox{under the form}\qquad \Lambda_{(k)}(s) =  \lambda (s)^k,  \qquad \Lambda(s) = \frac 1 {1- \lambda(s)}.
  \end{equation}
A Markov chain associated to the finite  alphabet $\Sigma$, is defined by the vector $R$  of initial  probabilities  $(r_i)_{i \in \Sigma}$ together with the  transition matrix $P:= [(p_{i| j})_{(i, j)  \in \Sigma\times \Sigma}]$. We denote by $P(s)$ the matrix with general coefficient $p_{i|j}^s$, and by $R(s)$  the vector of components $r_i^s$.  Then
\begin{equation} \label{lambda2}
\Lambda(s) = 1 + ^t \! \! {\bf 1} \cdot (I - P(s))^{-1} \cdot R(s).
\end{equation} If, moreover, the matrix $P$   is   irreducible and  aperiodic,  then, for any real $s$, the matrix $P(s)$  has a unique dominant eigenvalue $\lambda(s)$. \\
 
In both cases, the entropy  satisfies $ h({\cal S}) = -\lambda'(1)$.

\medskip {\bf 1.3. The first main  data structure:  the trie.}  A trie is a tree structure  which is used as  a dictionary in  various applications, as partial match queries,  text processing tasks or compression. This  justifies considering the trie structure
 as one of the central general purpose data structures of Computer Science. See  \cite{GoBa91} or 
 \cite{Sedgewick98b} for an algorithmic  study of this structure.

 \smallskip The trie structure  compares words  via their prefixes:  it is  based on a splitting according to symbols encountered. If ${\cal X}$ is a set of (infinite) words over  $\Sigma$, then the trie associated  to ${\cal X}$ is defined recursively by the rule:
 ${\tt Trie}({\cal X})$ is an internal node where are attached the tries $ {\tt Trie}({\cal X}\setminus a_1), {\tt Trie}({\cal X}\setminus a_2),\dots,{\tt Trie}({\cal X}\setminus a_r)$. Here, the set ${\cal X}\setminus a$ denotes  the subset of ${\cal X}$ consisting of strings that start with the symbol $a$
 stripped of their initial symbol $a$; recursion is halted as soon as ${\cal X}$ contains less than two elements:  if ${\cal X}$ is empty, then  ${\tt Trie}({\cal X})$ is empty; if  ${\cal X}$ has only one element $X$, then  ${\tt Trie}({\cal X})$ is a leaf labelled with  $X$.

 For $|{\cal X}| = n$, the trie ${\tt Trie}\, ({\cal X})$ has exactly $n$
 branches,  and the depth of a branch is the number of  (internal) nodes that it contains. 
 The {\em path--length} equals the sum of the depth of all branches: this is the  total number of symbols
that need to be examined in order to distinguish all elements of
${\cal X}$. Divided by the number of elements, it is also by definition
the cost of a positive search (i.e. searching for a word that is
present in the trie).  The {\em size\/} of the tree is the number of
its internal nodes.  Adding to the size,  the cardinality of ${\cal X}$ gives
the number of  prefixes necessary to isolate all elements
of ${\cal X}$. It gives also a precise estimate of the place needed in
memory to store
the trie in a real-life implementation. In this paper, we focus on two trie parameters: the size and the path-length. 

\vskip-0.3cm 
\paragraph{1.4. The second main data structure: the binary search tree ({\tt Bst}).}  We revisit here this well-known structure. Usually, this kind of tree contains keys and the path length of this tree measures the number of key comparisons that are needed to build the tree, and sort the keys by a method closely related to {\tt QuickSort}. This  usual cost    --the number of key comparisons-- is not realistic when the keys  have a complex structure, in the context of data bases or natural languages, for instance.  In this case, it is  more convenient to  view a key as a word, and now,  the cost for comparing two words (in the lexicographic order) is closely related to the length of their largest common prefix, called the coincidence.    The convenient  cost of  the bst is then  the total number of symbol comparisons between words that are needed to build it;  this is a kind of a weighted path length, called the {\em symbol path--length}   of the {\tt Bst}, also equal to the total symbol cost of {\tt QuickSort}.    For instance,  for inserting the key $F$ in the {\tt Bst} of Figure 1,  the number  of key  comparisons equals 3,  whereas the number of symbol comparisons equals  18 (7 for comparing  $F$ to $A$, 8 for comparing $F$  to $B$ and 1 for comparing  $F$ to $C$). This is this  symbol  path length that is studied in the following. 

 \begin{figure} [h]
\begin{tabular}{cc}
\begin{minipage}{10cm}
\vskip -1cm
\includegraphics[width= 9cm]{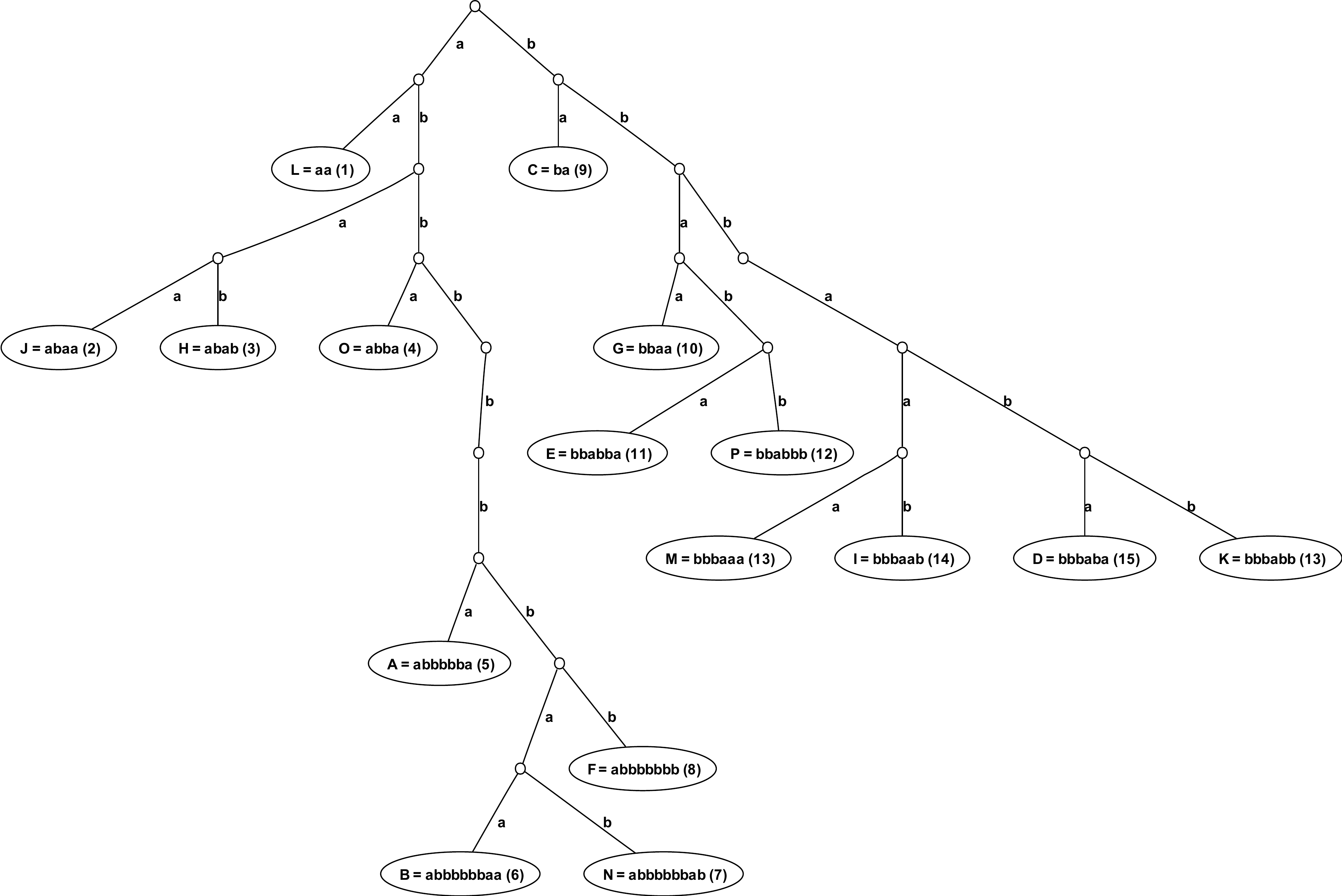}
\end{minipage} &  \begin{minipage}{8cm} \hskip -1cm\includegraphics[width= 7cm]{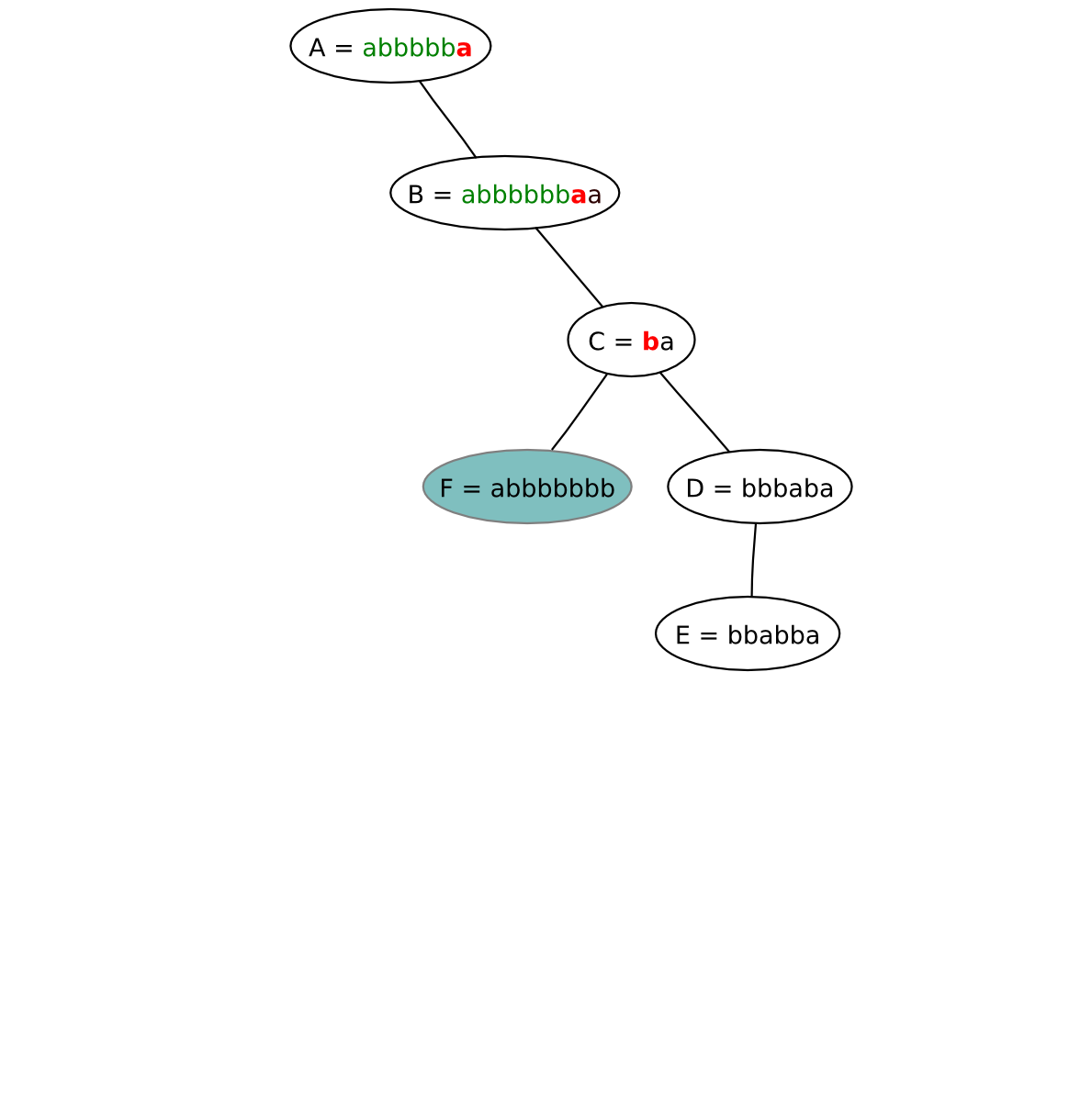}
\end{minipage}
\end{tabular}
\vskip-0.5cm 
\caption{\small On the left, a trie  built on  sixteen words of  $\{a, b\}^\star$. On the right, a  binary search tree built on  seven words  of   $\{a, b\}^\star$.}
\end{figure}

\medskip{\bf 1.5.  Average-case analysis:  exact expressions of the  three mean  costs.}
The average--case analysis of structures  (or algorithms) aims   characterizing
the mean value of their parameters under a well-defined probabilistic
model that describes the initial distribution of its inputs. Here, we
adopt the following quite general model: we work with a finite
sequence  ${\cal X}$ of infinite words independently produced by the same general 
 source 
 ${\cal S}$,  and we wish to estimate the mean value of the parameters when the cardinality $n$ of ${\cal X}$ becomes large.  Here, in the paper, we focus on three main parameters, two for  {\tt Trie}$({\cal X})$ and one for {\tt Bst}$({\cal X})$.  When restricted to simple sources,  there exist  many  works that study the trie parameters  (see \cite{Flajolet06, JaSz91, Knuth98a, Szpankowski01}) or the symbol path length for {\tt Bst} (see \cite{FiJa04}). The  same studies, in the case of a general source,    are  done  in  \cite{ClFlVa01} for the {\tt Trie} and in \cite{VaClFiFl09} for the {\tt Bst}, and are summarized as follows: 

 {\Thm {\rm [Cl\'ement, Fill, Flajolet, Vall\'ee]}. 
Let ${\cal S}$ be a   general source. Consider a finite sequence ${\cal X}$  of $n$   infinite words independently produced by ${\cal S}$.   Then the expectations   of  the  size  $R$ of {\tt Trie}$({\cal X})$,   the path length  $C$ of {\tt Trie}$({\cal X})$, the  symbol path--length $B$ of the binary search tree  {\tt Bst}$({\cal X})$ are   all expressed under  the  form 
\begin{equation} \label{Tn} T(n) = \sum_{k = 2}^n  (-1)^k  {n \choose k}  \varpi_T(k), 
\end{equation}
where the function $\varpi_T(s)$ is a Dirichlet series which  depends on  the parameter $T$ and  is closely related to the Dirichlet series $\Lambda(s)$ of the source ${\cal S}$,  defined in (\ref{Lambdadef})
\begin{equation} \label{varpi}
\varpi_R(s) =   (s-1)\,  \Lambda(s),   \qquad 
\varpi_{C} (s)  = s\,  \Lambda(s),   \qquad  \varpi_B(s) =    \frac {2}{s(s-1)}�\,  \Lambda(s). 
\end{equation}
}

 This result provides exact expressions for the mean values of parameters of interest,  that are totally explicit for simple sources, due to formulae given in (\ref{lambda1}) or in (\ref{lambda2}).    As we  now wish to obtain  an asymptotic form for these mean values, these nice exact   expressions are not easy to deal with, due to the presence of the alternate sum. The Rice formula, described in  \cite{Norlund29,Norlund54} and  introduced by Flajolet and Sedgewick \cite{FlSe95}  into the  analytic combinatorics domain,   transforms  an alternate sum into an integral of the complex plane, provided that the sequence of numerical values~$\varpi(k)$ lifts into
an analytic function~$\varpi(s)$. 
  
 {\em
Let $T(n) $ be a numerical sequence  which can be written as  in (\ref{Tn}), 
  where   the  function
$\varpi_T(s)$ is analytic   in  $\Re(s)> C$,  with $ 1<C <2$, and is there  
of polynomial growth  with order at most
$r$. Then  the sequence $T(n)$ admits a
\emph{N\"orlund--Rice representation},   for $n>r+1$ and   any   $C< d<2$. 
\begin{equation}\label{rice}
T(n)= \frac{1}{2i\pi}\int_{-d-i\infty}^{-d+i\infty}  \varpi_T(-s) \frac{ n!}{s(s+1)\cdots (s+n)}\, ds  
\end{equation}
}

\medskip{\bf 1.6. Importance of tameness of sources.}
The idea is now to push the  contour of integration in~\eqref{rice} to the right, past~$-1$.   This is why   we  
consider  the possible behaviours for the function $\varpi_T(s)$ near $\Re s = 1$,  more precisely on the left of the line $\Re s = 1$.   Due to the close relations between the functions $\varpi_T(s)$ and  the Dirichlet  series $\Lambda(s)$ of the source given in (\ref{varpi}), it is sufficient to consider  possible behaviours for $\Lambda(s)$ itself.  We  will  later show  why the   behaviours   that are described in the following definition, already given in  \cite{VaClFiFl09}\footnote{There are slight differences between the two definitions but the ``spirit'' is the same.}, and shown in Figure 2,   arise in a natural way for a large class of sources.  

 {\Def      Let  ${\cal R}$  be a region that contains the  half-plane $\Re s \ge 1$.   \\ A  source ${\cal S}$   is   ${\cal R}$--entropic  if   $\Lambda(s)$  is meromorphic   on ${\cal R}$ with  a simple    pole 
at~$s=1$,  simple, whose residue  involves the entropy   $h({\cal S})$ under the form  $1/h({\cal S})$. 

 A source is ${\cal R}$--tame  if  
 $(i)$  it is ${\cal R}$--entropic,  --
 $(ii)$ $\Lambda(s)$ has  no other pole than $s = 1$ in ${\cal R}$,  -- 
 $(iii)$  $\Lambda(s)$  is of polynomial growth  in ${\cal R}$ as  $|s|\to+\infty$.

    A source  is \\
 $(a)$  strongly--tame (S--tame in shorthand)  of abscissa $\delta$  if   there exists a vertical strip ${\cal R}$ of the form  $\Re(s)>1- \delta$, with  $ \delta >0$, where  $\Lambda(s)$ is ${\cal R}$-- tame.\\   
$(b)$  hyperbolically tame   (H--tame in shorthand) of exponent $\alpha$   if  there exists a  hyperbolic region  
 ${\cal R}$,  with $A, B, \alpha >0$ 
 $$ {\cal R} := \left\{ s = \sigma +it ;  \ \   |t| \ge B, \ \  \sigma> 1- \frac {A}{t ^\alpha}\right\} \bigcup \left\{ s = \sigma +it ;  \ \   \sigma>
 1- \frac {A}{B^\alpha}, |t| \le B \right\},  
  $$
 where $\Lambda(s)$ is ${\cal R}$--tame.   \\
 A source ${\cal S}$ is  periodic  of abscissa $\delta$, if    there exists a vertical strip ${\cal R}$ of the form  $\Re(s)>1- \delta$, with  $ \delta >0$, where  $\Lambda(s)$ is ${\cal R}$ entropic    and 
 admits a singularity  at a point  $1 +it_0$, for some real $t_0 >0$  
\footnote{This implies  that  $\varpi(s)$ admits singularities at all the points $1+ ik t_0$ for any integer $k$, and  is of polynomial growth on a family of  horizontal lines $ t = t_k$ with $t_k \to \infty$,   and  on vertical lines
$\Re(s)= 1- \delta'$ with  some $\delta'< \delta$}. }

For an entropic source, the Dirichlet  series $\varpi_T(s)$ has a pole of order 0 (for the {\tt Trie} size, cost $R$),  a pole of order 1 (for the {\tt Trie} path length, cost $C$),  a pole of order 2 (for the  {\tt Bst} symbol path length, cost $B$).

\begin{figure} [h]
 \begin{center}
 \includegraphics[width= 4cm]{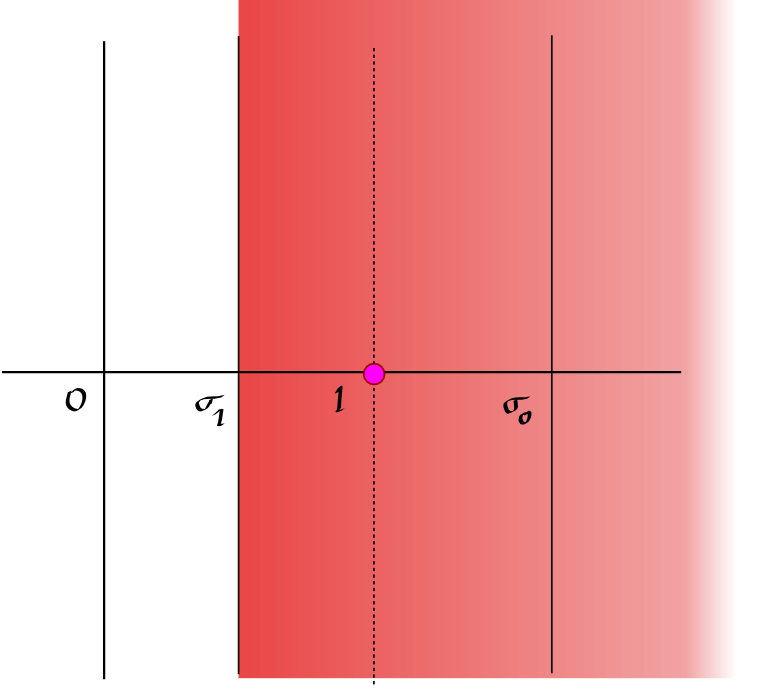} \includegraphics[width = 4.3cm] {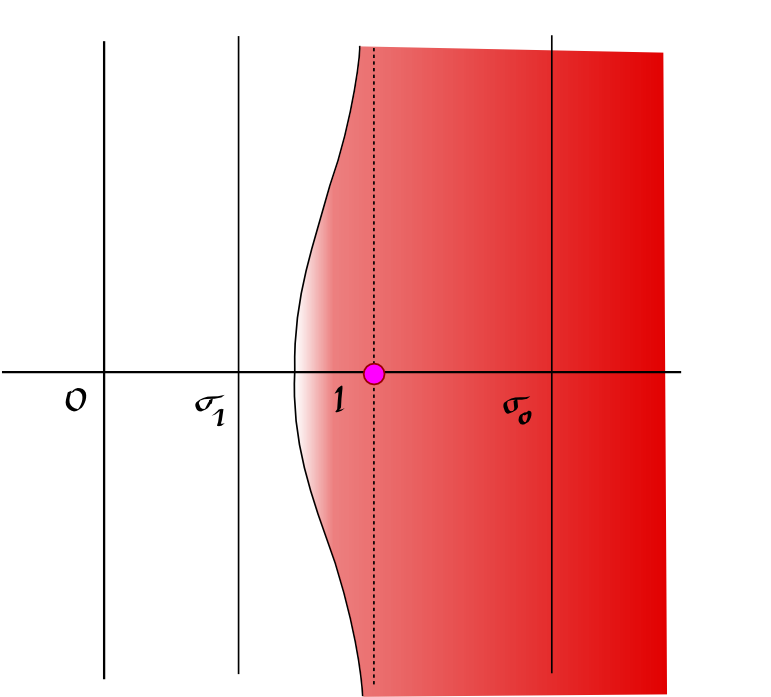} \includegraphics[width = 4.3cm] {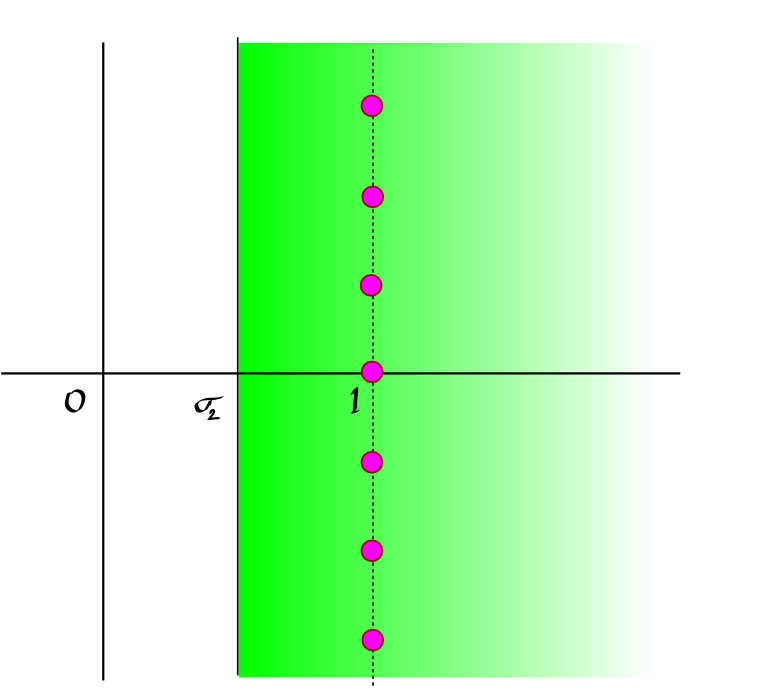}
 \end{center}
\caption{\small  Three possible domains  where the function $\varpi(s)$ is analytic and of polynomial growth.}
\end{figure}

\medskip {\bf 1.7. Average-case analysis:   asymptotic  expressions of the  three mean  costs.} 
 Now,  the following result shows that the  shape of the  tameness region  (described by the order, the abscissa, the exponent)  
essentially determine the  behaviour  of  the Rice integral in (\ref{rice}), and thus the asymptotic behaviour of our main parameters of interest:  It provides  a dictionary which transfers the  tameness properties of  the source  into asymptotic properties of the  sequence $T(n)$. The  following theorem gathers  and makes more precise results  that  are already obtained in   \cite{ClFlVa01} or \cite{VaClFiFl09} : 
 
{\Thm {\rm [Cl\'ement, Fill, Flajolet, Vall\'ee]}.  The  asymptotics of  each 
cost  $T(n)$ of  interest, relative to a parameter of a tree built on a  general source ${\cal S}$, and defined in Theorem 1,  
is of the   general following form 
$ T(n) =    P_T(n)  + E(n). $
 The ``principal term'' $P_T(n)$  involves the entropy $h({\cal S})$ under the form 
 $$ P_R(n) = \frac 1 {h({\cal S})} n,  \qquad  P_C(n) =    \frac 1 {h({\cal S})} n \log n  +a n, \qquad P_B(n) =   \frac 1 {h({\cal S})} n \log^2 n + b n \log n + cn,   $$
 together with some other constants $a,  b, c$.  
  The ``error term'' $E(n)$ admits the possible  following forms, depending on the tameness of the source

$(a)$ If ${\cal S}$ is  S--tame with  abscissa $\delta_0$,  then $E(n) =  O(n^{1-\delta})$, for any $\delta<\delta_0$.

$(b)$ If ${\cal S}$ is  H--tame with  exponent $\alpha_0$, then  $E(n) =  n \cdot O( \exp [-(\log n)^\alpha])$  for any     $\alpha < 1/(\alpha_0 +1)$.

$(c)$ If ${\cal S}$ is periodic  with   abscissa $\delta_0$, then  $E(n) =   n \cdot  \Phi ( \log n) +  O(n^{1-\delta}), $  for any $\delta<\delta_0$,\\
 where  $n \cdot \Phi(\log n)$ is the part of the expansion brought by the  family of the  non  real poles   located  on the vertical line $\Re s = 1$, and involves a periodic function $\Phi$.   
 }

Note that the ``error term'' $E(n)$  is not always ... an  actual error term: in the case of the trie size, for a periodic source, the fluctuation terms given by $E(n)$ arise in the main term. However, in all the other cases, the term $E(n)$ is indeed an error term.  The main term of the principal term always involves a constant equal to $1/h({\cal S})$, and the order of the main term depends on the tree parameter: it is always of the form $ n \log^k n$,  and the integer $k$ equals the order of the pole $s= 1$ for the Dirichlet series $\varpi_T(s)$ : one has  $k= 0$ for the {\tt Trie} size, $k = 1$ for the {\tt Trie} path length, and $k = 2$ for the {\tt Bst} symbol path length.  This result proves that, with respect to the number of symbol comparisons, the {\tt Bst} is  much less efficient  than the {\tt Trie}.

\medskip{\bf 1.8. Tameness of simple sources.}    We show  that   tameness properties  that are  described  in Definition 1 arise in a natural way  for  simple   sources.  Even if   S--tameness  never occurs for simple sources, we will see later that it   ``often'' occurs  for most of more ``complex'' sources.    We now focus on the  memoryless case,  defined by  the probabilities ${\frak P} = (p_1, p_2, \ldots, p_r)$,    to which we associate  
the ratios  $\alpha_{k, j} :=  {\log p_j}/{\log p_k}$. Then,  tameness properties  depend on arithmetic properties of the ratios  $\alpha_{k, j} $.%:=  {\log p_j}/{\log p_k}$. 

{\Pro Any  simple  source (memoryless source or irreducible aperiodic Markov chain)  is   entropic.  
A  memoryless source 
is  periodic 
if and only, for any fixed $k$,   all the real numbers  $\alpha_{k, j} $    are rationals with the same denominator}.

 We now  focus on  non-periodic memoryless sources, where there exists,  amongst all the reals 
 $\alpha_{k, j} $, 
  at least one  real $\alpha_{k, j}$ which is irrational. In this case,   there is no other pole of $\Lambda(s)$  than $s = 1$  on the vertical line $\Re s =  1$ but there exist poles  of $\Lambda$ which are arbitrary close to the vertical line $\Re s = 1$.   This entails that a  simple source is never strongly tame.  The distribution of   distances of the poles with respect to the vertical line $\Re s = 1$ depends on  the degree of approximability of the family $\alpha$ by rationals, as it was first remarked in \cite{FaFlHo86}.  We recall some notions on diophantine approximations (see  for instance \cite{Lagarias82b}).
  The irrationality exponent of a real $x$ is defined by
  $$ \omega(x) := \sup \left\{ \alpha, \left| x- \frac p q \right| \le \frac 1 { q ^{2+ \alpha}} \quad \hbox { for an infinite number of pairs  $(p, q)\in {\mathbb N}^2$} \right\}. $$
  A number  $x$  is diophantine  if its irrationality exponent is finite.   The following result provides a    characterisation  of  H--tameness for simple sources. It can be found in a more precise form in \cite{FlRoVa10}, where the authors revisit  previous results  of \cite{Lapidus06}. 
 
 {\Thm {\rm [Flajolet, Roux, Vall\'ee]}.   A memoryless source   is H--tame if and only  it is diophantine. 
  Moreover, there is a relation between the exponent  $\alpha$  of H--tameness   and the irrationality exponent $\mu(\frak P)$: one can choose as $\alpha$ any real strictly greater than $2 \mu(\frak P) +2$, and it is in a precise sense the best possible choice. }

  \medskip
  With the general Theorem 2, together with  Propositions 1 and 2, we can precisely describe   the asymptotic probabilistic  behaviour of two main  tree data structures  built on words produced by memoryless  sources. Generally speaking, Theorem 2 can be  applied   to tree structures built on  a general source as soon as  its tameness  may be studied. The following of the paper describes a general class of sources, which contains the simple sources,   for which tameness properties can be precisely studied.  We will see that tameness of these general sources may be quite different  from tameness of simple sources.

\section{Tameness of dynamical sources.}

We first define  the class of dynamical sources and  explain their relation with simple sources.   Then, we  recall the expression of the Dirichlet series $\Lambda(s)$ as a function of the secant transfer operator of the  underlying dynamical systems. Finally, we exhibit sufficient conditions on the underlying dynamical system under which it is possible to prove tameness properties [Theorem 4 for S--tameness, and Theorem 5 for H--tameness].

\medskip{\bf 2.1. Definition of dynamical sources.} A dynamical source, defined in \cite{Vallee01} is  closely related to  a dynamical system on the interval.

{\Def A dynamical system
of the interval  ${\cal I}:=[0,1]$  is defined by a mapping $T:{\cal I}\to {\cal I}$ (called the shift)  for which

$(a)$ there exists a  finite   alphabet  $\Sigma$,
and  a topological partition of
${\cal I}$ with disjoint open intervals $({\cal I}_m)_{m\in \Sigma}$, i.e.
$ {\cal I}=\bigcup_{m\in \Sigma}\overline{\cal  I}_m$.

$(b)$ The restriction of $T$ to each ${\cal I}_m$ is a ${\cal C}^2$
bijection from ${\cal I}_m$ to $T({\cal I}_m)$.

\pn The system is  complete when each restriction is surjective, i.e.,  $T({\cal I}_m) =
{\cal I}$.  The system is   Markovian  when  each  interval
$T({\cal I}_m)$ is a  union of intervals ${\cal I}_j$.
}

\smallskip
A dynamical system,  together with  a distribution $G$ on the unit interval ${\cal I}$,
defines a probabilistic source, which is called a dynamical source and is now described (See also Fig.1 at the end). 
The map $T$ is used as a shift mapping, and  the  mapping $\tau$
 whose restriction to
 each  ${\cal I}_m$ is equal to $m$,
 is used for coding.  The words are emitted  as follows [see Figure 3]:  To each real $x$, (except for a denumerable set),
  one associates   the trajectory 
  ${\cal T} (x) = (x, T (x), T^2 (x), \ldots T^j(x),  \dots)$,  which gives rise, via the  mapping $\tau$ to the word $M(x)\in\Sigma^{\mathbb N}$, 
  
\vskip 0.1cm 
 \centerline {$M(x)=(m_1(x),m_2(x),\dots,m_n(x),\dots) \qquad \hbox{with} \quad m_j(x) = \tau (T^{j-1}(x)).$}

 Given a prefix $w\in \Sigma^\star$,  the  set ${\cal I}_w$  of all  reals $x$  for which the word $M(x)$ begins with  the prefix $w$ is  an
interval, 
  the  fundamental interval associated to $w$, and the measure of this interval (with respect to distribution $G$),  is the fundamental probability $p_w$ of the source.
In the case of a complete system,  one denotes by  $h_{[m]}$ the local inverse of
$T$ restricted to ${\cal I}_m$ and  by $\cal H$ the set ${\cal H:=\{h_{[m]},
m\in {\Sigma}\}}$  of all local inverses.   Each  local inverse of the $k$--th iterate $T^k$  is  then associated to   a word $w =  m_1 m_2\ldots m_k \in \Sigma^k$; it is of  of the form
 $ h_{[w]}:= h_{[m_1]}\circ h_{[m_2]}\dots h_{[m_k]} $,  and 
   \begin{equation} \label{pw} {\cal I}_w =  h_{[w]} ({\cal I}), \qquad 
 p_w = |G(h_{[w]}(1))- G ( h_{[w]}(0))|.
 \end{equation}
  The set  of all the inverse branches of $T^k$ is   
   ${\cal H}^k = \{ h_{[w]}; \ \ w \in \Sigma^k\}$.  For
$h\in{\cal H}^k$, the number $k$ is called  the \emph{depth} of $h$
and it is denoted by $p(h)$.  We denote by  ${\cal H}^{\star} :=\cup_{k \ge 0} {\cal H}^k $ the set of all inverse branches.

\medskip
\pn Such sources may possess a high degree of correlations, due to the {\sl geometry} of the branches  and also to the {\sl shape} of branches.  

\pn  The geometry of the branches is defined by the respective positions of  ``horizontal'' intervals ${\cal I}_m$  with respect to ``vertical'' intervals   ${\cal J}_\ell := T({\cal I}_\ell)$ and 
allows to describe the   set ${\cal S}_m$  formed with   symbols  which can be possibly emitted after 
symbol $m$. The geometry of the system   then provides  a first access to the correlation between  successive symbols.  In particular, in a  {\em complete} system,   any  symbol of $\Sigma$  can be emitted 
after   any symbol $m$, and thus the equality ${\cal S}_m = \Sigma$ always holds.  

\smallskip
\pn The shape of the branches, and more precisely, the behavior of derivatives $|h'_m|$ has also a great   influence on correlations between symbols.  For a fixed geometry of the branches, a system with affine branches is ``less correlated'' than the other systems with the same geometry.  The contraction properties of ${\cal H}$, (i.e., the fact that $|h'_m|<1$) are  also essential, since they give rise to chaotic behaviour of the trajectories.

\medskip {\bf 2.2. Simple  sources  viewed as dynamical sources.} 
 All memoryless  sources and all Markov
chain sources belong to the general framework of  dynamical sources and 
 correspond to a piecewise linear shift, under this angle of dynamical sources. 
For
instance,  the  standard  binary system is  obtained  by $T(x)=\{2x\}$
($\{{\cdot}\}$  is  the fractional   part). 
 More precisely:  
 
 -- A  memoryless source is a  complete dynamical source, with  affine branches and a uniform  initial distribution, \\
  --   A  Markov chain is a Markovian dynamical source, with affine branches and   a family of uniform  initial distributions on each ${\cal J}_j$.

%\end{itemize}
Figure 3 shows  three instances of  simple sources, viewed as dynamical sources.

However,  as soon as the derivatives $h'$ of the branches  are not constant, there  exist  correlations between successive   symbols, and the dynamical source  is no longer  simple.   Dynamical sources with a non-linear shift allow for
correlations   that  depend  on the   entire    past.
A main instance is the dynamical source relative to the Gauss map,  represented in Figure 3, which underlies the Euclid Algorithm and is defined on the unit interval via the shift $T$
\begin{equation} \label{T}
 T(0) = 0, \qquad T(x) = \frac 1 x -  \left \lfloor \frac 1 x  \right \rfloor  \quad (x \not = 0).
 \end{equation}

 \begin{figure}[h]
 \begin{center}
  \begin{tabular}{cc}
 \begin{minipage}{4.5cm}
 \includegraphics[width =4cm]{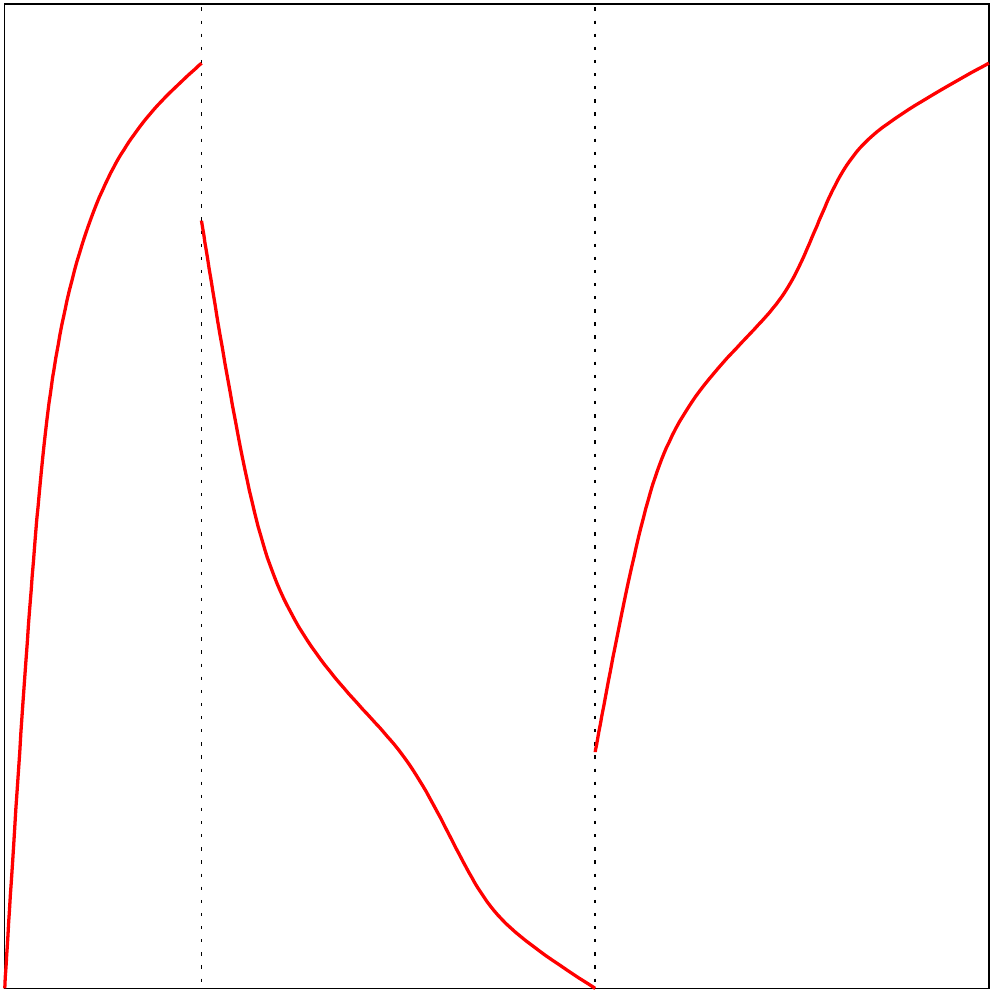}  
 \end{minipage} &
  \begin{minipage}{4.5cm}
  \includegraphics[width = 4cm]{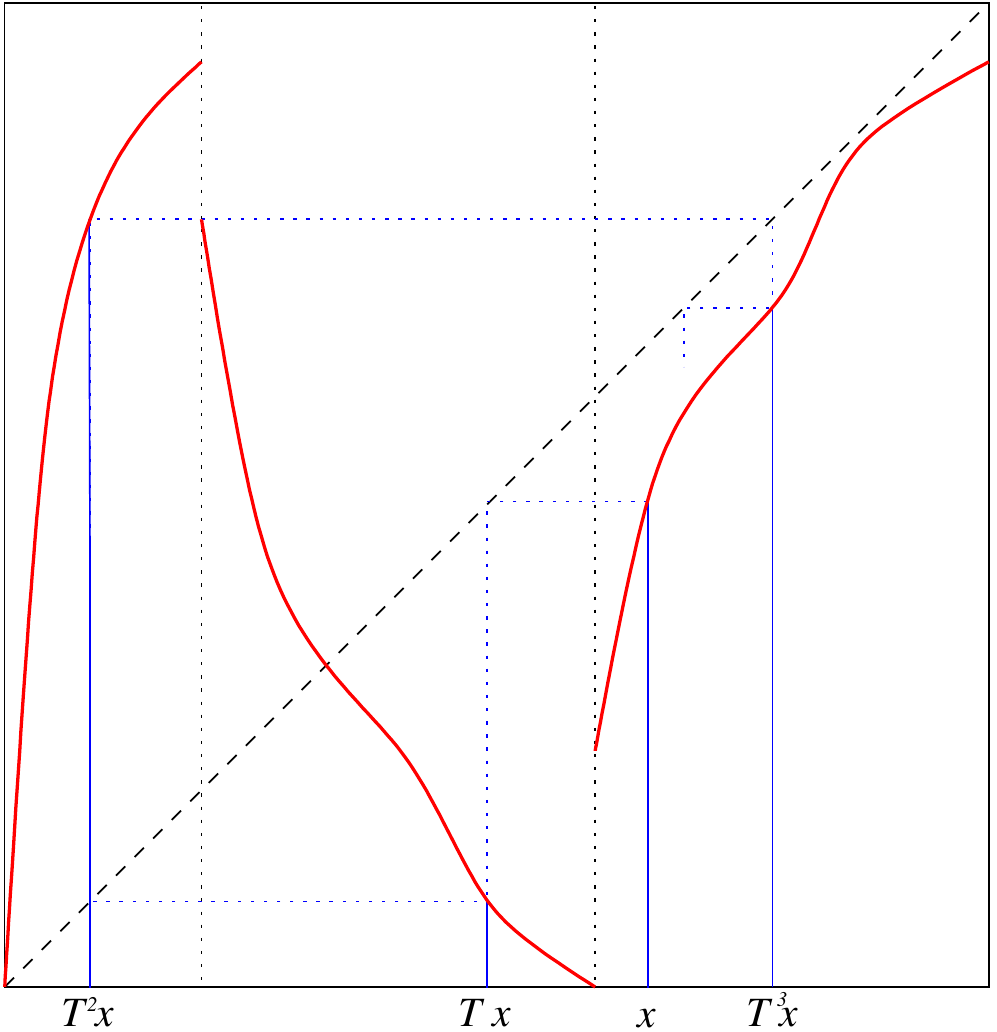}
  \end{minipage}
  \end{tabular}
   \end{center}
     
  \vskip 0.5cm 
 \begin{tabular}{cccc}
 \begin{minipage}{3.8cm}
 \includegraphics[width=3.5cm]{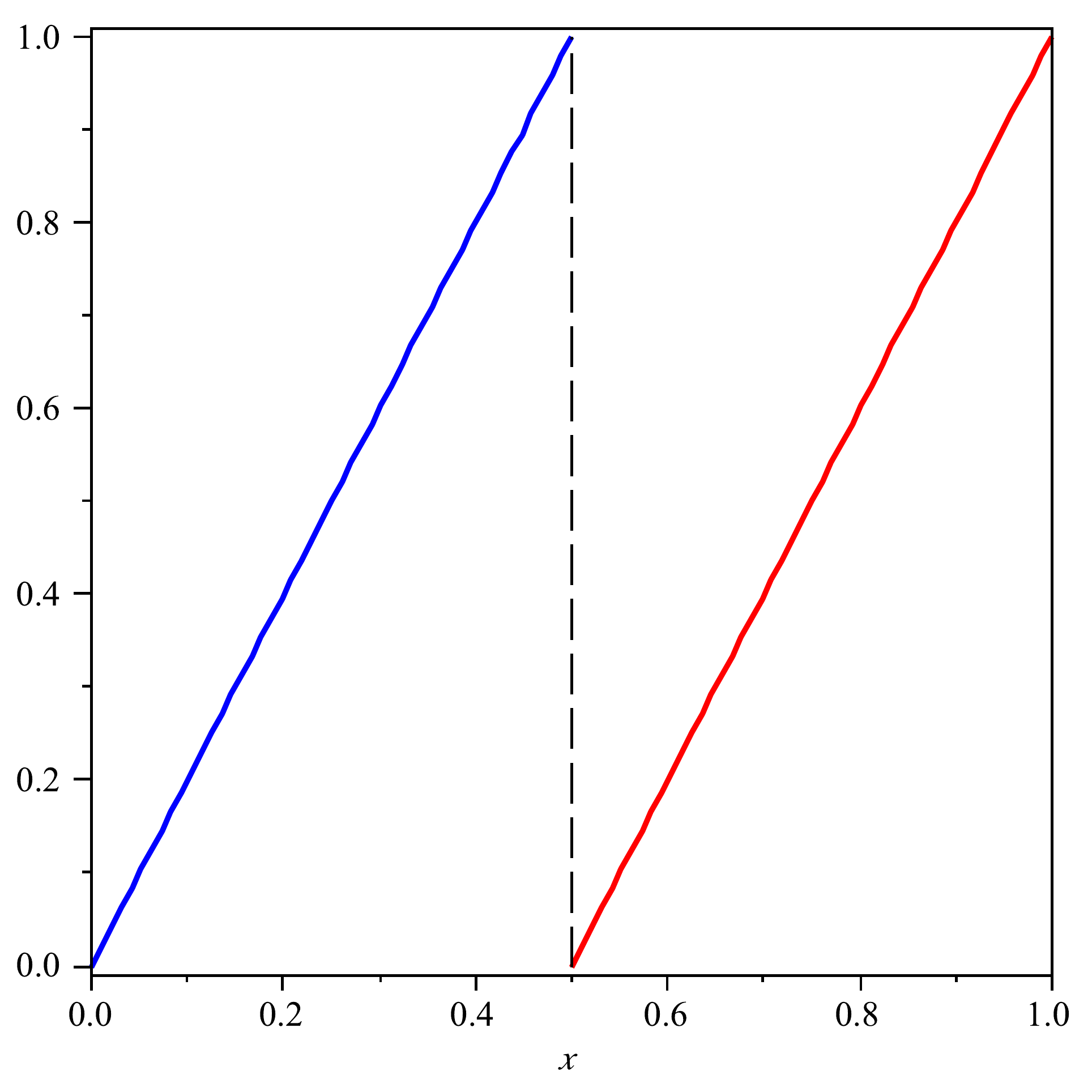}\ 
 \end{minipage} &
 \begin{minipage}{3.8cm}
\includegraphics[width=3.5cm]{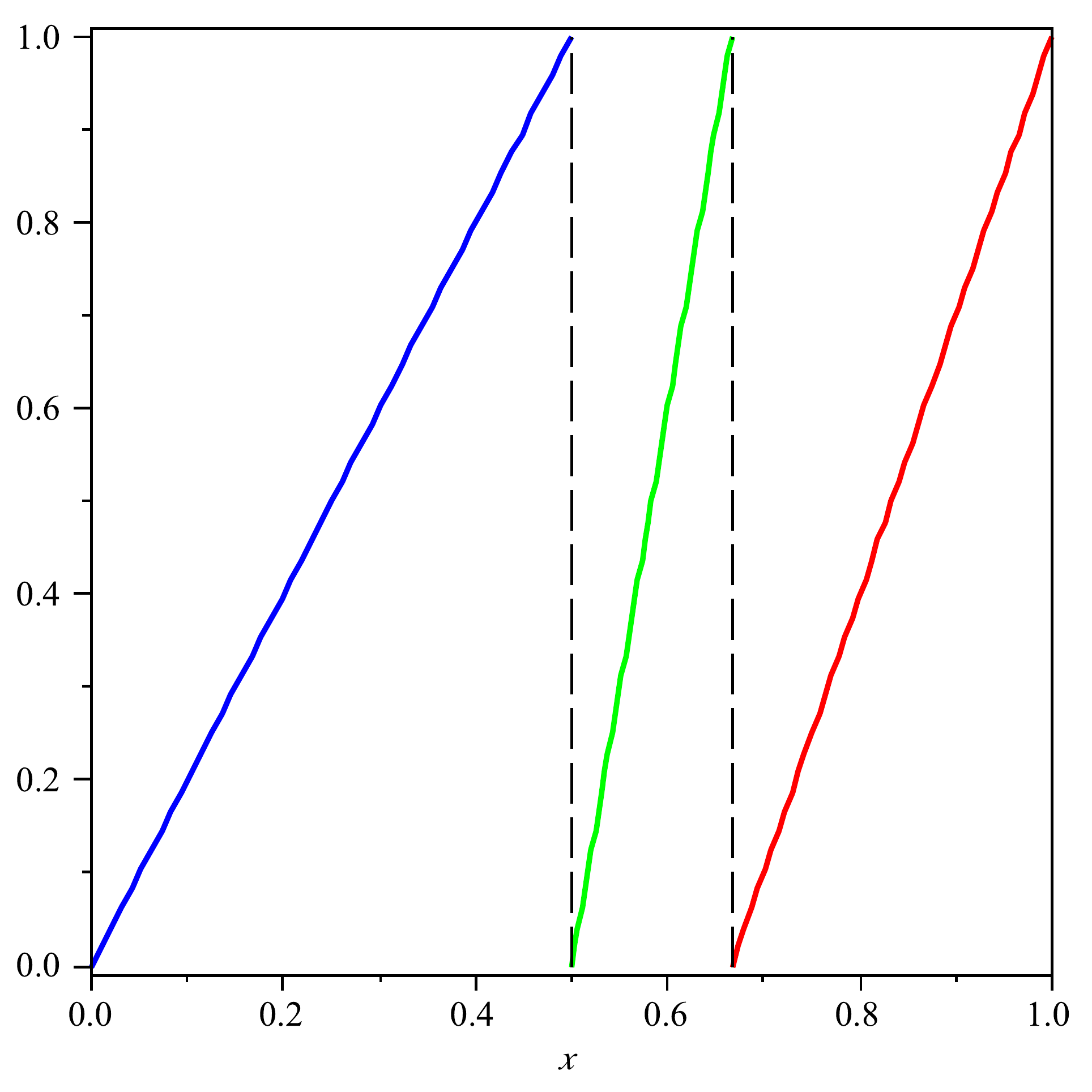}  
\end{minipage}&
 \begin{minipage}{3.8cm}\vskip -0.1cm
\includegraphics[width=3.1cm]{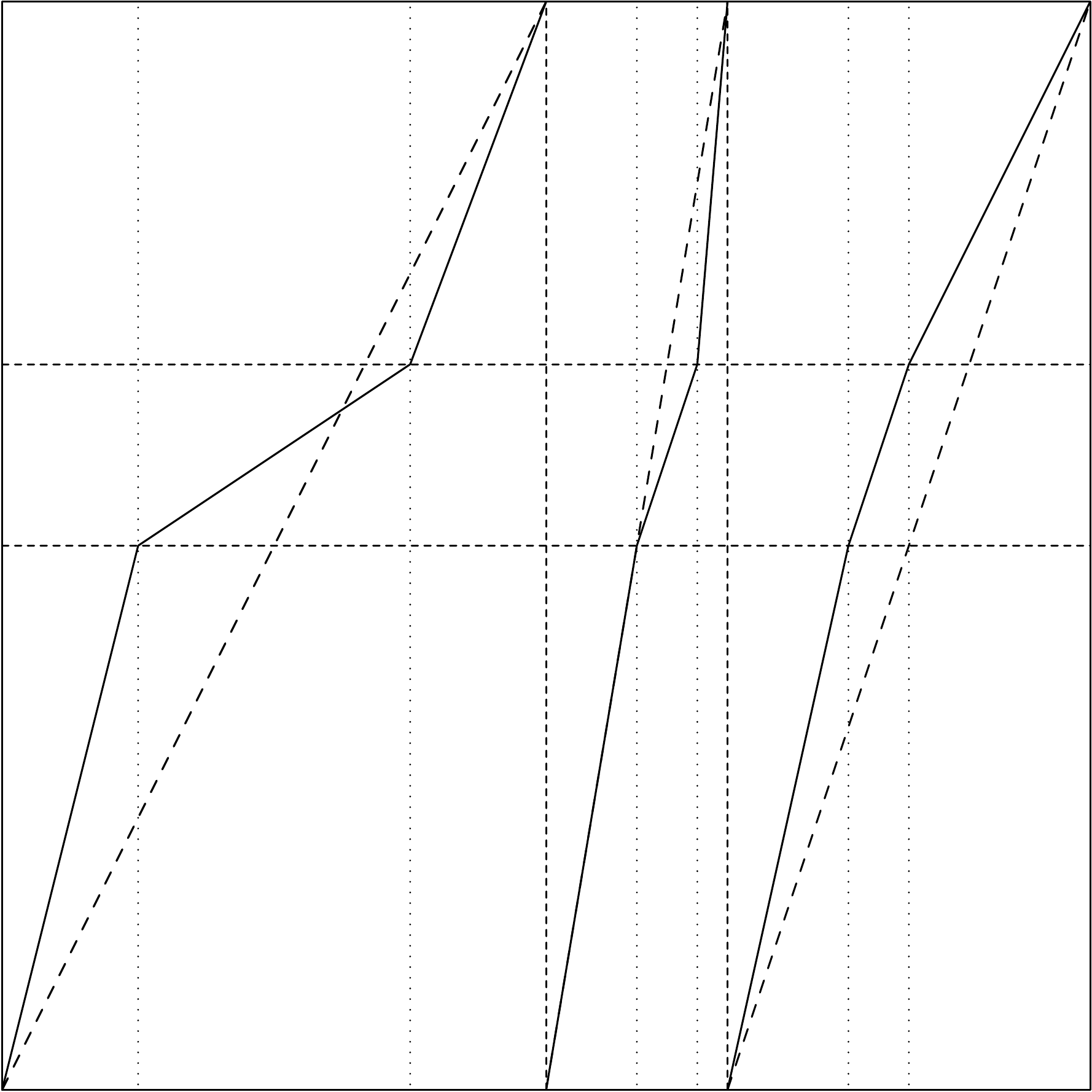} 
\end{minipage} &
\begin{minipage}{4cm} \hskip -0.7cm \includegraphics[width =4cm]{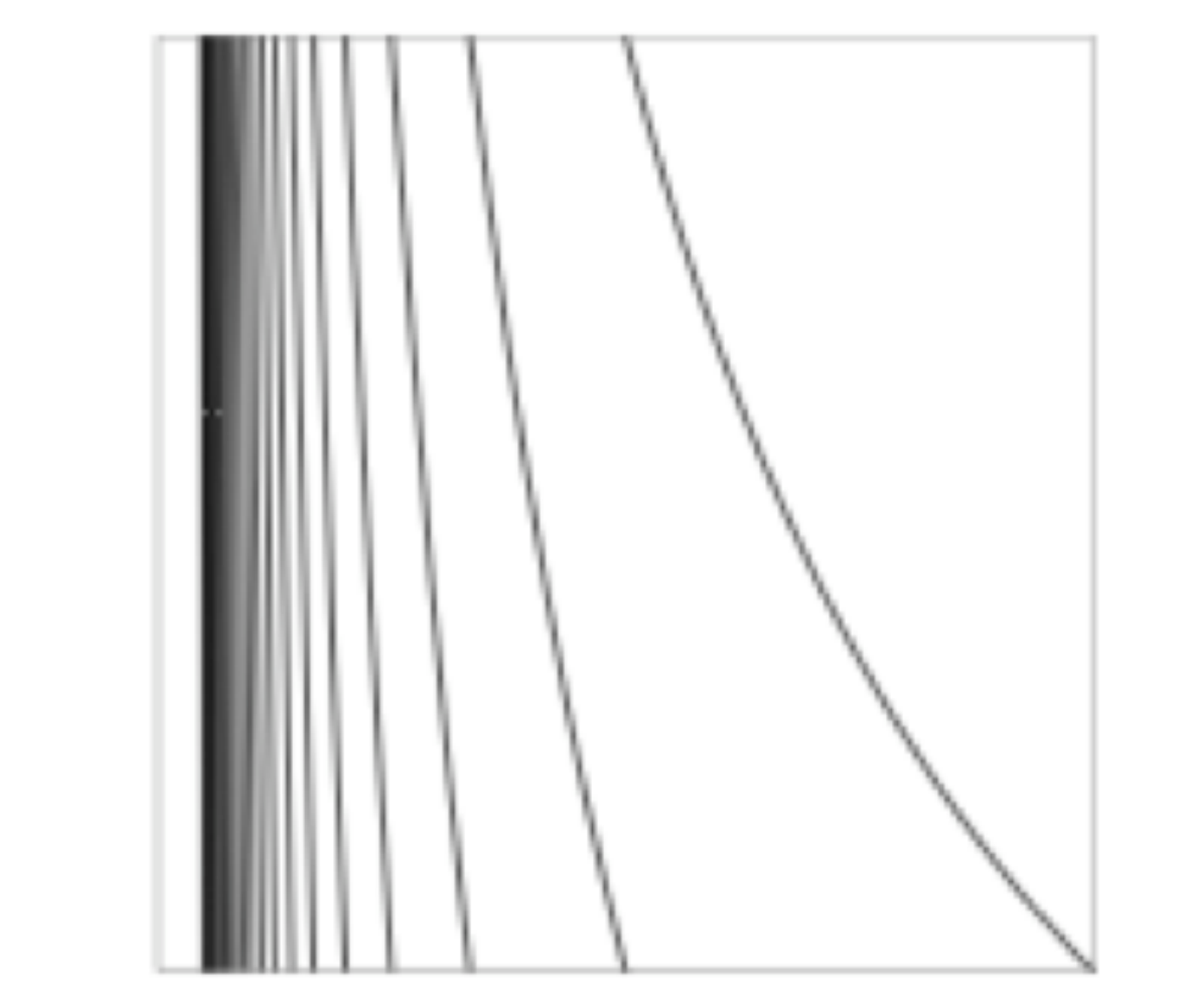}  
\end{minipage}
\end{tabular}

\caption{\small    (Up)  A dynamical system, with  $\Sigma= \{a, b, c\}$ and a  word  $M(x) = (c,b, a, c \ldots)$ --  (Down) Two memoryless sources     and a Markov chain, viewed as dynamical sources. The continued fraction source.}

\end{figure}

\medskip
{\bf 2.3. Transfer operators.}
One of the main tools  in dynamical  system theory is the transfer operator
introduced by Ruelle, denoted by $\H_s$. It generalizes
the density transformer $\H$ that describes the evolution of the
density. 

\pn We  here consider the case of a complete dynamical system: if $f=f_0$ denotes the initial density on $\cal I$, and
$f_1$ the density on $\cal I$ after one iteration of $T$, then $f_1$
can be written as $f_1=\H[f_0]$, where $\H$ is defined by
$$ \H := \sum_{h \in \HH}  \H_{(h)}  \qquad \hbox {with}\quad  \H_{(h)} [f](x):= |h'(x)| \, f\circ h(x).$$
The transfer operator extends the density transformer; it
  depends on a complex parameter $s$, 
 \begin{equation}\label{Hs}
{  \H}_s= \sum_{h\in{\cal H}} \H_{(h), s}   \qquad \hbox {with}\quad  \H_{(h), s} [f](x):= |h'(x)|^s \cdot f\circ h(x),
\end{equation}
and coincides with $\H$ when $s = 1$. Here, we are interested by  generating the fundamental probabilities,  whose expression is provided in (\ref{pw}) in the case of a complete dynamical system. The main tool is a generalized version of the transfer operator --the secant  transfer operator-- introduced  by Vall\'ee in \cite{Vallee01}. 
 This operator  involves the secant function
of inverse branches (instead of their derivatives), it acts on functions $F$ of two variables; for $s \in {\mathbb C}$,  and $h \in {\cal H}$,  we first  define  the  component  secant operator  ${\bH}_{(h), s}$  as
\begin{equation}
\bH_{(h), s}
[F](x,y):=
\left|\frac{h(x)-h(y)}{x-y}\right|^sF(h(x),h(y)), 
\label{secant0}
\end{equation}
\vskip -0.5cm
 \begin{equation}
 \hbox {and  the  secant transfer operator   is defined as} \qquad \bH_s :=  \sum_{h\in\HH}\  \bH_{ (h), s}, \hskip 4.5cm 
\label{secant}
\end{equation}
 Denote by  ${\rm diag}\,  F$ 
 the function  defined by ${\rm diag}\,  F(x) := F(x, x)$. The equality 
$ \bH_s[F](x, x) = \H_s[{\rm diag\, } F](x)$
 holds on the diagonal $x = y$  and shows that  the secant operator is   an extension of the plain transfer operator. Moreover, 
multiplicative properties of  secants then entail the relation
$$   \bH ^k_s =   \sum_{h \in {\cal H}^k}  \bH_{(h), s} \qquad \hbox{so that} \quad {  \bH}^k_s[F](x, y)= \sum_{h \in {\cal H}^k}
\left|\frac{h(x)-h(y)}{x-y}\right|^sF(h(x),h(y)) .$$
Finally, 
 the Dirichlet series  can be expressed as a quasi--inverse of the secant operator: this is a nice extension of the expressions  obtained for simple sources, in (\ref{lambda1},\ref{lambda2}). 
  
 {\Pro{\rm [Vall\'ee]}.    For a  complete  dynamical source, relative to a shift $T$ and a distribution $G$,
 the  Dirichlet series of the source admits  an alternative expression  which involves  the quasi--inverse of the secant operator,
  defined in (\ref{secant}) 
 applied to the function $L^s$, where $L$ is  the secant of the distribution $G$,  $$  \Lambda (s) = (I-  {\bH}_{s})^{-1}  [L^{s}](0, 1), \qquad \hbox{with} \quad  L(x, y) :=  \frac { G(x)- G(y)} {x-y} .$$
}

\smallskip {\bf 2.4. Tameness of dynamical sources.}   Here, we consider  subclasses of dynamical sources, for which   the quasi-inverse has nice spectral properties. This will entail, with Proposition  2,  nice properties for the function $\Lambda(s)$, from which one deduces  tameness properties.  The main results are as follows: 
There exist  natural instances of dynamical  sources which  are    S--tame,  or H--tame.  A  ``random'' dynamical source  is ``very often'' S--tame: this happens as soon as   its inverse branches  have  ``not too often''   the same ``shape''.   A  dynamical source  can be  periodic only if it  ``closely resembles'' a memoryless source.   A dynamical source is H--tame if,   informally speaking, its arithmetical properties are the same as the arithmetical properties of a  H-tame memoryless source.  More precisely, we define three (large) subclasses of dynamical sources -- the {\tt Good} Class, the {\tt UNI} Class, the {\tt DIOP} Class-- for which we can  describe  the tameness in an informal setting.   The {\tt UNI} Class has been already studied and described in previous works \cite{dolgo1,BaVa05,BaVa05b, CeVa10}. The original part of our  work  is related to the {\tt DIOP} Class, for which we revisit and extend previous results  described in \cite{dolgo2, Mel, Na}.   We  first state the main tameness results for dynamical sources  in an informal way: 

\medskip   
{\bf Theorem.}   {\em All the sources of the {\tt Good-UNI} Class  are S--tame. 
  All the sources  of the {\tt Good-DIOP} Class  are H--tame.   A source of the {\tt Good} Class may be  periodic  only if   it is  conjugated to  a source with affine branches. }

\medskip{\bf 2.5. The {\tt Good} Class.}
 We first define  the 
{\tt Good} Class, for which the shift is expansive, and gives rise to a chaotic behaviour for the trajectories.  

 {\Def 
\label{dss} {\rm [{\tt Good} Class]}. A dynamical system
of the interval   $({\cal I}, T)$  belongs to the {\tt Good} Class if   it is complete, with a set ${\cal H}$ of inverse branches which satisfies the following:

 $(G1)$ The set ${\cal H}$ is uniformly contracting, i.e., there exists   a constant  $\rho<1$,   for which
$$   \forall h \in {\cal H},   \quad   \beta_h:= \sup\{  |h'(x)|; \ \ x \in {\cal I} \}  \le \rho.
$$

$(G2)$  There is a constant $A>0$,
    so that every inverse branch $h\in {\cal H}$ satisfies
     $|h''|\le A |h'|$.

$(G3)$ There exists $\sigma_0 <1$ for which the series
    $\sum_{h\in {\cal H}}\beta_h^{s}$ converges on $\Re s>\sigma_0$.

}
  The essential condition is $(G1)$. The bounded distortion property $(G2)$  and the  property $(G3)$ are technical conditions that 
 always fulfilled   for a finite  alphabet $\Sigma$.

When the dynamical system  belongs to the {\tt Good} Class,  the transfer  operators   (tangent and secant) act on spaces of functions of ${\cal C}^1$ class. They admit dominant spectral
 properties  for  $s$ near the real axis, together with a spectral gap. This implies that, for $s$ near 1,
  the function $\Lambda(s)$   is  meromorphic for $s$ with a small imaginary part, and  admits a simple p\^ole at $s = 1$.

\medskip{\bf 2.6.  The  {\tt UNI} Condition. }
 One   first defines a probability  $\Pr_n$ on each set  ${\cal H}^n \times {\cal H}^n$, in a natural way, and     lets  $\Pr_n \{(h, k)\} := | h({\cal I})|\cdot  | k({\cal I})|, $ where  $|{\cal J}|$ denotes the length of the interval ${\cal J}$.  Furthermore,
 $\Delta (h, k)$ denotes
the ``distance''
 between  two inverse branches $h$ and $k$  of same depth,  defined as
\begin{equation} \label {delta}
\Delta (h, k) = \inf_{x \in {\cal I}} |\Psi'_{h, k} (x)|  \qquad
\hbox{with} \quad   \Psi_{h, k} (x)= \log \left|\frac
{h'(x)}{k'(x)}\right|.
\end{equation}
The distance $\Delta(h, k)$ is a measure of the difference between the ``shape'' of  the two branches $h, k$.
The {\tt UNI} Condition, stated   as follows \cite{dolgo1},  is a geometric condition which expresses that  the  probability that two inverse branches   have almost the same ``shape'' is  very  small: 

{\Def  {\rm  [Condition {\tt UNI}]}.     A dynamical system $({\cal I}, T)$ satisfies the {\tt UNI} condition   if its  set ${\cal H}$ of inverse branches satisfies  the following

 $(U1)$ For
any $a \in ]0, 1[$,  and for any integer $n$, one has $\Pr_n [ \ \Delta \le \rho^{an}]   <\!\!< \rho^{an}$.

$(U2)$ Each $h \in {\cal H}$ is of class ${\cal C}^3$ and  for any $n$,   there exists $B_n$ for which $|h'''| \le B_n |h'|$ for any $h \in {\cal H}^n$.}

 For a source with affine branches, the ``distance''  $\Delta$ is always zero, and the probabilities of Assertion $(U1)$ are all equal to 1.  Such a source  never satisfies the Condition {\tt UNI}.  Conversely,
  a  dynamical source of the {\tt Good-UNI}  Class  cannot be  conjugated  to a source with affine branches, as it is proven  by Baladi and Vall\'ee \cite{BaVa05}. Then, the condition {\tt UNI}  excludes all the simple sources, which cannot be S--tame. The strength of the  Condition  {\tt UNI} is due to the fact that this condition  is sufficient to imply  strong tameness :

  {\Thm  {\rm [Dolgopyat, Baladi--Vall\'ee, Cesaratto--Vall\'ee]}  When the dynamical system  of the {\tt Good} Class satisfies the condition {\tt UNI},  it gives rise to a  S--tame source.}
   
  There are natural instances of sources that belong to the {\tt Good-UNI} Class,  for instance    the Euclidean dynamical system defined in (\ref{T}),  together with two other  dynamical  systems, of the Euclidean type.

  \medskip{\bf 2.7.  The diophantine conditions.}  The {\tt Good-UNI} Class gathers systems  which are quite different from systems with affine branches. The {\tt DIOP} Condition ``copies''  the behaviour  of memoryless  sources, when they are H--tame. In this case,  we recall that there exists a  ratio
  $  {\log  p_i}/{\log p_k}$
  which is   diophantine,  i.e.,   whose irrationality exponent is finite.

 The {\tt DIOP} condition  is an arithmetical condition, which  extends this condition to a system of the Good Class.   For an inverse branch $h$, one denotes by  $h^\star$   its  unique  fixed point (such a point exists and is unique for a system of the Good Class),   by  $p(h)$  its depth,  and one lets,   for  $h, k, \ell$ in  ${\cal H}^\star$, 
   $$ 
    c(h) = \frac  {\log |h'(h^\star)|} {p(h)}, \qquad c(h, k) = \frac {c(h)}{c(k)},  \qquad c(h, k, \ell) = 
 \frac {c(h)-c(k)}{c(h)-c(\ell)}.$$
 We can now state the definition of diophantine dynamical sources:

 {\Def {\rm [{\tt DIO2} and  {\tt DIOP3}]}.  A dynamical source is  2--diophantine ({\tt [DIOP2]} in shorthand) if there exist two branches  $h$ et $k$ of   ${\cal H}^\star$  for which the ratio $c(h, k)$  is diophantine.\\
  A dynamical source is  3--diophantine ({\tt [DIOP3]} in shorthand) if there exist three  branches  $h$,  $k$  and $\ell$ of   ${\cal H}^\star$  for which the ratio $c(h, k, \ell)$  is diophantine}
  
 The following result proves that  these conditions are sufficient to entail H--tameness of associated sources. This is the main   contribution of  Roux' PhD thesis \cite {Ro}.  The appendix contains hints on the proof, that  will be detailed in the long version.

{\Thm {\rm [Dologopyat, Naud, Melbourne, Roux--Vall\'ee] } \\
$(a)$ A dynamical system of the {\tt Good} Class,  which  is moreover {\tt DIOP3},  gives rise to a H--tame source.  \\
$(b)$ A dynamical system of the Good Class,  which  is moreover {\tt DIOP2},  gives rise to a H--tame source.}

\medskip{\bf 2.8.  A little piece of history.}  Dolgopyat, in two seminal papers \cite{dolgo1,dolgo2},   introduces  the  Conditions {\tt UNI}  and {\tt DIOP2}.  He proves that, under these conditions, the quasi-inverse of the plain (tangent) transfer operator  has nice properties  in a region on the left of the line $\{\Re s = 1\} $:  when the {\tt UNI} Condition  holds, the region is a vertical strip, and when the {\tt DIOP2} Condition holds, the region is of hyperbolic type. However, he does not consider the case of an infinite number of branches, and  his results are extended to this  case by Baladi and Vall\'ee in \cite{BaVa05,BaVa05b}  for the {\tt UNI} condition, and by Melbourne \cite{Mel} in the case of the {\tt DIOP} condition, who introduces the {\tt DIOP3} Condition. However, in order to  deal with the   Dirichlet series $\Lambda(s)$, one needs to extend the previous proofs  to the secant operator. This have been done by Cesaratto and Vall\'ee in \cite{CeVa10} for the {\tt UNI} Condition. Here, we deal with the {\tt DIOP}  conditions and  we perform two extensions: we consider a possible infinite alphabet,  we  deal both with  the {\tt DIOP3} (where we use a method due to Melbourne \cite{Mel})  and the {\tt DIOP2} condition (where we use a method due to Naud \cite{Na}). We  also extend these results to the secant operator.

\medskip
{\bf Acknowledgements.}    This work takes place inside  the ANR project {\tt MAGNUM} [M\'ethodes Algorithmiques pour la G\'en\'eration Non Uniforme: Mod\`eles et Applications] [ANR 2010 BLAN 0204] .

\section{Some hints on the proof of Theorem 5.}

Since the Dirichlet series $\Lambda(s)$ is expressed with the quasi-inverse of the secant operator $\bH_s$ 
we study  the behaviour of this quasi-inverse on the vertical line $\Re s = 1$.  It is closely related to the behaviour of the  operators $ {\bf M}_t, \bM_t$ defined by 
$$  {\bf M}_t [f](x) := |T'(x)|^{it} f\circ T(x),  \qquad   {\bM}_t [F[(x, y) := \left|\frac {T(x-T(y)}{x-y}\right|^{it}  F( T (x), T(y)) .$$

\medskip {\bf 3.1. Various possibilities for the spectral radius of the operator $\bH_s$ on  $\Re s = 1$.}  The beginning point is the following proposition,  that is classical for the  tangent operator,  and  can be easily extended to the secant operator.

\medskip
{\bf Proposition 3.} {\sl   Consider a dynamical system of the {\tt Good} Class  and its secant transfer operator $\bH_s$, acting on the  space ${\cal C}^1({\cal I} \times {\cal I})$  for a parameter  $s$ of the form $s = 1+it_0$,  with   $t_0 \not = 0$. 

$(a)$  For a   complex number  $\lambda$ of modulus 1,  the two conditions are equivalent:

 \qquad$(a1)$   The complex number $\lambda$  belongs to the  spectrum  ${\rm Sp \, }  {\bH}_{1+ it_0}$ .
 
 \qquad$(a2)$   The complex number $\lambda^{-1}$   is an eigenvalue  of $  {\bf M}_{t_0}$.

\medskip
 $(b)$  Assume that there exists $t_0 \not = 0$  for which the condition  $(a2)$  is satisfied.    
   Then,  there exist  $a\not = 0$ and  $b$   for which the quantities  $c(h)-  b$  all belong to the  ${\mathbb Z}$-module  ${\mathbb Z}a$.
 
 \qquad $(b1)$   If  $\lambda $ is a root of unity,    then  
 all the ratios  $c(h, k)$  are rationals.

\qquad $(b2)$    If $\lambda$ is any complex number of modulus 1,  all the ratios $c(h, k, \ell)$  are rationals.

\medskip
$(c)$   If one of the two conditions is satisfied
  
 \qquad   $(c1)$  there exists a ratio   $c(h, k)$  which is not rational,

 \qquad$(c2)$
  For any  $t \not = 0$,   the spectrum of the operator  ${\bM}_t$   does not contain  $\lambda = 1$. 

  then, the quasi-inverse $(I- {\bH}_{s})^{-1}$  is analytic on  $\Re s = 1$  except at $s = 1$  where it has a simple pole.

\medskip 
$(d)$   If one of the two conditions is satisfied
  
 \qquad$(d1)$  there exists a  ratio  $c(h, k, \ell)$ which is not rational,

 \qquad $(d2)$  For any $t \not = 0$,   the spectrum of the operator  ${\bM}_t$   does not contain any  $\lambda$ with $|\lambda| = 1$, 
  
   then, the spectral radius of  ${\bH}_s$  is strictly less than  1 on  $\{s; \Re s = 1, s \not = 1 \}$  and,   for any  $\lambda$  of modulus 1,    the quasi-inverse  $(I- \lambda {\bH}_{s})^{-1}$ is analytic on the line  $\Re s = 1$  except at $s = 1$  where it admits a simple pole. 
     }

\medskip{\bf 3.2. Reinforcement of  conditions $(c), (d)$. } 
    The main question is now as follows:  if one of the conditions $(c1)$ or  $(d1)$ or $(c2)$  or $(d2)$ is replaced by a stronger condition, is it possible to obtain  a conclusion about tameness, of the following  kind:

{\sl $(R)$    There exists a region on the left of the vertical line $\Re s = 1$  on which the quasi-inverse $(I- {\bH}_{s})^{-1} $  is analytic except  at $s= 1$ (where it admits a simple pole), and is of polynomial growth for $t = \Im s \to \infty$. }

 We deal here  with the Banach space ${\cal C}^1 ({\cal I} \times {\cal I})$ formed with functions of class ${\cal C}^1$ on the unit square, endowed with the norm $||.||_1$ defined by $||u||_1 := \sup  |u(x, y|  +\sup  ||u'(x, y)|| $, but we  also use a norm $||.||_{(t)}$ which depends on the imaginary part  $t$ of $s$, defined by  $||u||_{(t)} := \sup |u(x, y|  + (1/|t|) \sup ||u'(x, y)||$. Our main object of study is
 \begin{equation} \label{Rt}  {\cal R}( t) :=  
 \left \| \left(I -   {\bH}_{1 +it}\right)^{-1} \right\| _{(t)}.  
 \end{equation}
 
\medskip
A  possible reinforcement  {\tt DIOP3}  of the condition  $(d1)$   is ``There exists a triple  $(h, k, \ell)$  for which  $c(h, k, \ell)$  is diophantine''. A possible  reinforcement ($d3$)  of the condition $(d2)$  is:   ``The operator  ${\bM}_t$  does not admit a system of  almost eigenfunctions''  for which a more formal statement will be provided later.  We will also see that these two reinforcements are not independent since the implication   {\tt DIOP3} $\Rightarrow$ ($d3$) holds

A  possible reinforcement   {\tt DIOP2}  of the condition  $(c1)$   is ``There exists a pair  $(h, k)$  for which  $c(h, k)$  is diophantine''. A possible  reinforcement ($c3$)  of the condition $(c2)$  is:   ``The operator  ${\bM}_t$  does not admit a system of almost  invariant functions''  for which a more formal statement will be provided later. We will also see that these two reinforcements are not independent since the implication  {\tt DIOP2} $\Rightarrow$ ($c3$) holds

\paragraph{\bf 3.3. Precise statement of Theorem 5.} There  are two theorems, one for each condition {\tt  DIOP2} or {\tt  DIOP3}.  

\smallskip
{\bf Theorem 5$(a)$.} [DIOP3] { \sl  Consider a dynamical source of the {\tt Good} Class, with a possibly infinite denumerable  alphabet, with a contraction ratio  $\rho<1$.    If there exists a triple $\{h, k, \ell\}$, with $\nu = \max \{ c(h), c(k),$
$ c(\ell)\}$,   for which  $c(h, k, \ell)$ is diophantine with exponent  $\mu$,  then ${\cal R}(t)$  is of polynomial growth, with an exponent    strictly larger than  $$ 4\mu  +3  +  \nu \frac{2\mu +4}{ |\log \rho|}    .$$
}

{\bf Theorem 5$(b)$.} [DIOP2] {\sl Consider a dynamical source of the {\tt Good} Class, with a possibly infinite denumerable  alphabet, with a contraction ratio  $\rho<1$ and a pression function    $s \mapsto L(s)$\footnote{the pression is the logarithm of the dominant eigenvalue $\lambda(s)$}.    Consider the real $\nu_1$ defined from the pressure function by the two equations 

\centerline {  $ (1-\sigma_1)L'(\sigma_1)+ L(\sigma_1) = \log \rho, \ \  \nu_1 = -L'(\sigma_1)$. }
  If there exists a pair $\{h, k\}$,  with $\nu_0 = \max \{ c(h), c(k) \}$,   for which  $c(h, k)$  is diophantine with exponent  $\mu$,    then ${\cal R}(t)$  is of polynomial growth,  with an exponent  strictly larger than
$$  4\mu  +3  +  \nu \frac {2\mu +4} {  |\log \rho|}    \qquad  \hbox { with} \ \    \nu =  \max (\nu_0,\nu_1) .  $$
    }
  
\medskip
{\bf 3.4.  Main sets of interest.}   One considers triples $({\cal T}, {\cal W}, \eta)$   formed  with

\qquad $(i)$  a subset  ${\cal T}$  of the set $\{ t \in {\mathbb R}, |t| \ge 1 \}$, 

 \qquad $(ii)$  a family ${\cal W}$  of functions, \ \  
 ${\cal W} := \{w_t, \ t \in {\cal T}; \ \   w_t \in  {\cal C}^1({\cal I}\times {\cal I}),  \ \  |w_t| = 1,  \ \   ||w_t||_{(t)} \le K \}$, 
 
  \qquad $(iii)$  a family  $\eta$  of complex numbers, \ \  
 $\eta := \{ \eta_t \in {\mathbb C}, \ \ t\in {\cal T} , \ \  |\eta_t| = 1 \}.$

We consider properties which are satisfied only on  subsets of the unit  square  ${\cal I} \times{\cal I}$ and only in an approximative way, and, for a given imaginary part $t$, these  subsets, and the approximation will depend on $t$  (in a polynomial way), and there are various parameters  $(\alpha, \beta, \gamma, \delta)$ for the possible exponents.

One lets  $n(\beta, t) := \lceil \beta \log |t| \rceil$, $n(\theta, t) := \lceil \theta \log |t| \rceil$  and   considers the following subsets of ${\cal H}^\star$, 
$$   {\cal H}(t, \theta) :=  {\cal H}^{n(\theta, t)}, \qquad {\cal H}(t, \beta, \delta) :=  \left\{ h \in {\cal H}^{n(\beta, t)}; \quad \min 
\{ |h'(x)|, \,  x \in {\cal I} \}\ge \frac 1 {t^\delta} \right\}.$$ 
The following subsets  of ${\cal I} \times {\cal I}$  are called ``fundamental  unions''
$$
  {\cal I}(t, \beta, \delta):=   \bigcup_{h \in {\cal H}(t, \beta, \delta)} h ({\cal I})\times h({\cal I}), \qquad   {\cal I}(t, \beta, \delta, \theta):=
   \bigcup_{h \in {\cal H}(t, \beta, \delta)
   \atop {\ell \in {\cal H}(t, \theta)}} h\circ \ell  ({\cal I})\times h\circ \ell ({\cal I}) .$$
  %The last subset is called a ``fundamental  union''. 

\smallskip
In the proof, there are  various subsets which intervene: Subsets ${\cal A}$, related to the notion of ``almost eigenfunctions'' --
subsets ${\cal C}$ relared to the notion   of ``almost  invariant functions'' -- subsets ${\cal E}$ which approximate  subsets ${\cal A}\setminus {\cal C}$ -- subsets ${\cal B}$ related to the behaviour  of the iterate  of the  operator --  Subsets ${\cal F}$ related to the growth of the quasi-inverse of the secant operator.  The final subset of interest is the subset ${\cal F}$, and the other ones form a chain of subsets which will be compared to ${\cal F}$ in the proof.  The first three ones involve the approximate subset ${\cal I}(t, \beta, \delta, \theta)$.

    The set   ${\cal A} (\alpha, \beta, \delta, \theta)$ gathers all the reals  $t$
  for which there exists a pair $(w_t, \eta_t)$   that satisfies,  
\begin{equation}
\label{Mt} | {\bM}_t^{n( \beta, t)} [w_t](x, y) - \eta_t w_t(x, y) | \le  \frac {1}{t^\alpha}, \qquad \hbox{for any  $(x, y) \in {\cal I}(t, \beta, \delta, \theta). $}
\end{equation}

\medskip
 The set  ${\cal C}( \alpha, \beta, \gamma, \delta, \theta, k_0)$  gathers all the reals  $t$  for which there exists a pair $(w_t,   \eta_t)$    that satisfies
\begin{equation}
\label{MtC}    |\eta_t^{k_0}- 1| \le  \frac 1 {t^\gamma}. \qquad | {\bM}_t^{n( \beta, t)} [w_t](x, y) - \eta_t w_t(x, y) | \le  \frac {1}{t^\alpha}, \qquad \hbox{for any $(x, y) \in {\cal I}(t, \beta, \delta, \theta). $}
\end{equation}

\medskip
 The set ${\cal E}( \alpha, \beta,  \gamma,  \delta, \theta, k_0)$  gathers the reals  $t$ for which there exists a pair $(w_t,  \eta_t)$   that satisfies
\begin{equation}
\label{MtE} |\eta_t^{k_0}-   1| >  \frac 1 {t^\gamma}, \qquad 
  | {\bM}_t^{n( \beta, t)} [w_t](x, y) - \eta_t w_t(x, y) | \le  \frac {1}{t^\alpha} ,  \qquad \hbox{for any  $(x, y) \in {\cal I}(t, \beta, \delta, \theta)$}. \end{equation}
 The inclusion   ${\cal A}(\alpha, \beta, \delta, \theta) \setminus {\cal C}(\alpha, \beta, \gamma,\delta, \theta, k_0) \subset {\cal E}( \alpha, \beta,  \gamma,  \delta, \theta, k_0)$ holds.

\medskip
Let  $\psi$ be the invariant function of ${\bH}_1$. 
 The set   ${\cal B} (\alpha, \beta, \theta)$     gathers all the reals  $t$   for which there exists $u_t$, with $||u_t||_{(t)}\le 1$,   that satisfies, for any  $n \le 3 n(\beta, t)$  and any $(x, y) \in  {\cal I}(t,  \theta) $, 
$$|  {\bH}^n _{1 + it}  [\psi u_t](x, y)| \ge  \psi(x, y) \left( 1- \frac {1}{t^\alpha}\right).$$

 The set  $ {\cal F}(\alpha)$   gathers the  reals $t$  for which the quasi-inverse   is of polynomial growth  with exponent $\alpha$
  $$  {\cal F}(\alpha) :=  \{ t, \ \   {\cal R}(t) \le  t^\alpha \}. $$
We wish to prove that there exists $\alpha$ for which ${\cal F}^c(\alpha)$  is bounded.

\medskip{\bf 3.5. Relation between diophantine properties, subsets  ${\cal A}$ and  ${\cal C}$.} 
There are two main results, described in Lemma 0 (with subset ${\cal A}$) and Lemma 1 (with subset ${\cal C}$). 

\medskip
{\bf Lemma 0. }
{\sl  Consider a triple  $(h, k, \ell) \in {\cal H}^3$  and a real  $\alpha >1$.     
 If there exists a triple   $(\beta, \delta, \theta)$   with 
\begin{equation}\label{betadelta}
 \frac{\delta}{\beta} > \max \{c(h), c(k), c(\ell)\}, \end{equation}   for which
 $ {\cal A}(\alpha, \beta, \delta, \theta) $  is unbounded, 
 then   $c(h, k, \ell)$ 
  has an irrationality exponent at least equal    to $ \alpha -1$. \\
 If  $c(h, k, \ell)$ is diophantine with exponent $\mu$,   then for any  4-uple
$( \alpha, \beta,  \delta, \theta)$ avec $\alpha > \mu +1$, et    $(\beta, \delta)$  which satisfies (\ref{betadelta}),   the subset 
$ {\cal A}(\alpha, \beta, \delta, \theta)$ is bounded. }

\bigskip
{\bf Lemma 1.}
{\sl  Consider a pair  $(h, k) \in {\cal H}^2$  and a real  $\mu >1$.     If there exists  a 6--uple $(\alpha, \beta, \gamma, \delta, \theta, k_0)$  with $\min (\alpha, \gamma) = \mu$,   and \begin{equation}\label{betadelta1}
 \frac{\delta}{\beta} > \max \{c(h), c(k) \}, \end{equation}
 for which the subset 
 $ {\cal C}(\alpha, \beta, \gamma,  \delta, \theta, k_0)$  is unbounded, 
 then   $c(h, k)$ 
    has an irrationality exponent at least equal    to $\mu -1$.  \\
If  $c(h, k)$ is diophantine with exponent $\mu$,    then, for each 6--uple  $(\alpha, \beta, \gamma, \delta, \theta, k_0)$  with $\min (\alpha, \gamma)  > \mu +1$ et $(\beta, \delta) $  that satisfies  (\ref{betadelta1}),  the subset  
$ {\cal C}(\alpha, \beta, \gamma,  \delta, \theta, k_0)$  is bounded. }

\smallskip
 In the following of the  proof, we use  the notion of weak inclusion   between two  subsets ${\cal L}$ et ${\cal M}$   de ${\mathbb R}$. The subset ${\cal L}$ is said to be weakly included in ${\cal M}$  [this is denoted by $ {\cal L}\inc  {\cal M}$] if there exists $t_1 \in {\mathbb R}$ for which 
${\cal L} \cap[t_1, + \infty[  \subset  {\cal M} \cap[t_1, + \infty[ .$

\medskip{\bf 3.6. Relations between subsets  ${\cal A}$ and ${\cal F}$.}  Lemma 2 compares  subsets ${\cal B}$ and ${\cal A}$ whereas Lemma 3 compares subsets ${\cal B}$ and ${\cal F}$. Lemmas 2 and 3 are summarized in Lemma 4 which compares subsets ${\cal A}$ and ${\cal F}$. Lemmas 0 and  4  together    prove Theorem 5$(a)$.

\medskip
{\bf  Lemma 2. } {\sl  
  For any 4-uple  $(\alpha, \beta, \delta, \theta)$  that satisfies $\beta |\log \rho| \ge \alpha +1$,  % let $\alpha_1 = 2 \alpha + \delta$. Then, 

   \centerline{the weak inclusion
 ${\cal B}(\alpha_1, \beta, \theta) \inc   {\cal A}( \alpha, \beta, \delta, \theta)$  holds, for any $\alpha_1 >2 \alpha+ \delta$.}
 }

 \smallskip
{\bf  Lemma 3. }{\sl  For any triple  $(\alpha_1, \beta, \theta)$ that satisfies $\theta |\log \rho| \ge \alpha_1 +1$,   

\centerline{
 the weak inclusion
 $ {\cal B}^c (  \alpha_1, \beta, \delta)   \inc   {\cal F}(\alpha_2) 
 $ holds for any $\alpha_2 > 2 \alpha_1 +1$}
 }

 \smallskip
 {\bf Lemma 4.}  {\sl     For any 4-uple $(\alpha, \beta, \delta, \theta)$  that satisfies \   
 $\beta |\log \rho| \ge  \alpha + 1, \quad \theta |\log \rho| \ge 2 \alpha + \delta +1, $
   
   \centerline{ 
 the weak inclusion $ {\cal F}^c( \alpha_2 )  \inc   {\cal A}( \alpha, \beta, \delta, \theta)$ holds, for any 
 $\alpha_2 >4 \alpha + 2 \delta +1$  }   }

\medskip {\bf 3.7. Relation  between subsets  ${\cal E}$  and ${\cal F}$.}    This relation is described in Lemma 5. 
   
  \medskip
   {\bf Lemma 5.} { \sl   One considers the logarithm  $L(s)$ of the dominant eigenvalue $\lambda(s)$  of   the operator $\bH_s$ and  the real $\nu_1$ defined from the pressure function by the two equations 

\centerline {  $ (1-\sigma_1)L'(\sigma_1)+ L(\sigma_1) = \log \rho, \ \  \nu_1 = -L'(\sigma_1)$. }  For any 5--uple $(\alpha, \beta, \gamma,  \delta, \theta)$   which satisfies the relations 
   $$ \alpha>\gamma ,  \qquad \beta \ge   \frac {\alpha +1} {|\log \rho|},  \qquad  \frac {\delta}{\beta} > \nu_1,$$  there exists an integer $k_0$ for which the weak inclusions 
   
   \centerline{
   ${\cal E}(\alpha, \beta,  \gamma, \delta, \theta, k_0) \inc   {\cal F}( 2 \alpha),  \qquad \hbox{and thus } \quad 
   {\cal F}^c( 2 \alpha)  \inc  {\cal A}^c(\alpha, \beta, \delta, \theta) \cup {\cal C}(\alpha, \beta, \gamma, \delta, \theta, k_0)$ hold.}}

   \medskip {\bf 3.8. Relation  between subsets  ${\cal C}$ and  ${\cal F}$.}  One gathers the conclusions of Lemmas 4 and 5 in Lemma 6.   Lemmas  1 et 6   together prove  Theorem 5$(b)$.  
   
\medskip    
   {\bf Lemma 6.} {\sl One considers the logarithm  $L(s)$ of the dominant eigenvalue $\lambda(s)$  of   the operator $\bH_s$ and  the real $\nu_1$ defined from the pressure function by the two equations 

\centerline {  $ (1-\sigma_1)L'(\sigma_1)+ L(\sigma_1) = \log \rho, \ \  \nu_1 = -L'(\sigma_1)$. } 
 For any 5--uple $(\alpha, \beta, \gamma,  \delta, \theta)$   which satisfies the relations 
   $$\ \alpha> \gamma ,   \qquad \beta \ge   \frac {\alpha +1} {|\log \rho|},  \qquad  \theta  \ge  \frac {2 \alpha+ \delta  +1} {|\log \rho|} ,   \qquad  \frac {\delta}{\beta} > \nu_1,$$     there exists an integer $k_0$ for which 
    the weak inclusion
    
   \centerline{  $  {\cal F}^c(  2\alpha + \delta)  \inc  {\cal C}(\alpha, \beta, \gamma, \delta, \theta, k_0) $ holds.}
   }

  \end{document}